\documentclass[conference]{IEEEtran}
\IEEEoverridecommandlockouts
\usepackage{diagbox} 
\usepackage{graphicx}
\usepackage{epstopdf}
\usepackage{psfrag}
\usepackage{subfigure}
\usepackage{url}
\usepackage{stfloats}
\usepackage{amsfonts,amssymb,amsmath,bm,paralist,theorem,cite,ifthen,color,nccmath}
\usepackage[font=normalsize,labelfont=bf]{caption}
\usepackage{calc}
\usepackage{enumerate}
\usepackage{multirow}
\usepackage{makecell}
\usepackage[ruled]{algorithm2e}
\usepackage{setspace}
\usepackage{array}
\usepackage{float}
\usepackage{soul}
\usepackage{bbm}
\usepackage{cases}
\usepackage{lipsum} 
\usepackage{etoolbox}
\usepackage[colorlinks,
            linkcolor=blue,
            anchorcolor=red,
            citecolor=red]{hyperref}
\usepackage[]{mdframed}

\SetKwInOut{Input}{Input}
\SetKwInOut{Output}{Output}

\newcommand{\T}{{\scriptscriptstyle\mathsf{T}}}
\renewcommand{\H}{{\scriptscriptstyle\mathsf{H}}}

\allowdisplaybreaks

\graphicspath{{Figures/}}

\newcommand\Ocl{\ensuremath{\mathcal{O}}}

\newcommand\Lcl{\ensuremath{\mathcal{L}}}

\newcommand\Cs{\ensuremath{{\mathbb{C}}}}

\newcommand\Fb{\ensuremath{ \mathbf{F} }}
\newcommand\Gb{\ensuremath{ \mathbf{G} }}
\newcommand\Hb{\ensuremath{ \mathbf{H} }}
\newcommand\Ib{\ensuremath{ \mathbf{I} }}

\newcommand\Wb{\ensuremath{ \mathbf{W} }}

\newcommand\Xb{\ensuremath{ \mathbf{X} }}
\newcommand\Yb{\ensuremath{ \mathbf{Y} }}

\newcommand\hb{\ensuremath{ \mathbf{h} }}

\newcommand\ssb{\ensuremath{ \mathbf{s} }}

\newcommand\wb{\ensuremath{ \mathbf{w} }}
\newcommand\xb{\ensuremath{ \mathbf{x} }}

\newcommand\zb{\ensuremath{ \mathbf{z} }}

\newcommand\Sigmab{\ensuremath{{\bm \Sigma}}}

\newcommand\diag{\ensuremath{{\rm diag}}}

\newcommand\Nt{\ensuremath{ N_{\rm t} }}
\newcommand\Nr{\ensuremath{ N_{\rm r} }}

\newcommand\Pt{\ensuremath{ P_{\rm t} }}

 \usepackage{afterpage}

\usepackage{etoolbox}
\newcommand{\zerodisplayskips}{%
  \setlength{\abovedisplayskip}{1pt}%
  \setlength{\belowdisplayskip}{1pt}%
  \setlength{\abovedisplayshortskip}{1pt}%
  \setlength{\belowdisplayshortskip}{1pt}}
\appto{\normalsize}{\zerodisplayskips}
\appto{\small}{\zerodisplayskips}
\appto{\footnotesize}{\zerodisplayskips}

\usepackage[font=footnotesize,labelfont=bf]{caption}

\usepackage{enumitem}

\newcommand{\eqnsize}{\small}

\title{Model-Based Machine Learning for Max-Min Fairness Beamforming Design in JCAS Systems \vspace{-0.5cm}}
\author{
\IEEEauthorblockN{Mengyuan Ma$^*$, Tianyu Fang$^*$, Nir Shlezinger$^\dagger$, A.~L.~Swindlehurst$^\S$, Markku Juntti$^*$, and Nhan Nguyen$^*$}
\IEEEauthorblockA{$^*$Centre for Wireless Communications (CWC), University of Oulu, Finland \\
$^\dagger$School of ECE, Ben-Gurion University of the Negev, Beer-Sheva, Israel \\
$^\S$Department of EECS, University of California, Irvine, CA, USA\\
Email: \{mengyuan.ma, tianyu.fang, markku.juntti, nhan.nguyen\}@oulu.fi; nirshl@bgu.ac.il; swindle@uci.edu \vspace{-0.5cm}
}}	

\begin{document}

\maketitle
\vspace{10mm}
\setlength{\abovedisplayskip}{3.5pt}
\setlength{\belowdisplayskip}{3.5pt}
\begin{abstract}
    Joint communications and sensing (JCAS) is expected to be a crucial technology for future wireless systems. This paper investigates beamforming design for a multi-user multi-target JCAS system to ensure fairness and balance between communications and sensing performance. We jointly optimize the transmit and receive beamformers to maximize the weighted sum of the minimum communications rate and sensing mutual information. The formulated problem is highly challenging due to its non-smooth and non-convex nature. To overcome the challenges, we reformulate the problem into an equivalent but more tractable form. We first solve this problem by alternating optimization (AO) and then propose a machine learning algorithm based on the AO approach. Numerical results show that our scheme scales effectively with the number of the communications users and provides better performance with shorter run time compared to conventional optimization approaches.
\end{abstract}


\section{Introduction}\label{sec:intro}

Joint communications and sensing (JCAS) is a pivotal technology for future wireless communications systems, enabling communications and sensing functionalities on a unified hardware platform. This integration facilitates spectrum sharing and reduces hardware costs \cite{zhang2018multibeam, ma2020joint, zhang2021overview}. However, the shared and limited resources for these dual functionalities pose significant challenges for JCAS transceiver design \cite{liu2022survey, demirhan2023integrated}. Beamforming plays a key role in addressing this challenge by enhancing spectral efficiency and improving sensing accuracy \cite{nguyen2023deep, ma2024joint}.


Recent literature on beamforming design for JCAS systems has focused on three main optimization goals: maximizing sensing performance, maximizing communications performance, and balancing the tradeoff between them. Sensing-oriented designs \cite{johnston2022mimo, Hassanien2016Dual, huang2020majorcom, liu2018mu, liu2020joint, qi2022hybrid, chen2024transmit, liu2021cramer, song2023intelligent, zhu2023cramer, lin2023target, liu2024bi, lin2024deep, chen2023joint,tang2018spectrally} optimize sensing while ensuring communication capabilities. In contrast, communication-oriented designs \cite{nguyen2023multiuser, choi2024joint, liu2022transmit} prioritize communications performance under sensing constraints. Other works \cite{liu2018toward, cheng2021hybrid, nguyen2023joint, wang2024joint, chen2021joint, wang2023optimizing, fang2024beamforming, zhang2024joint, elbir2021terahertz,peng2024mutual} explore the communications-sensing performance tradeoff by maximizing a weighted sum of communications and sensing utility functions. Communications performance is typically measured by rate or SINR, while sensing performance is evaluated using metrics like beampattern matching~\cite{liu2018mu, liu2020joint, johnston2022mimo, qi2022hybrid}, SCNR~\cite{choi2024joint, wang2024joint, wang2023optimizing}, sensing mutual information (MI) \cite{peng2024mutual,tang2018spectrally}, and the Cramér--Rao lower bound~\cite{liu2021cramer, song2023intelligent, zhu2023cramer}.

Despite these advances, achieving fairness among multiple communications users and sensing targets remains a significant challenge due to the inherently non-smooth nature of max-min optimization problems \cite{chen2021joint, wang2023optimizing, fang2024beamforming}. Conventional optimization-based approaches \cite{chen2021joint, wang2023optimizing, fang2024beamforming} typically result in high complexity and involve numerous algorithm parameters that need to be well tuned to ensure performance. In contrast, model-based machine learning (ML) methodologies~\cite{shlezinger2023model} facilitate real-time operation and satisfactory performance for beamforming design \cite{demirhan2023integrated, shlezinger2024artificial}. For example, beamforming optimizers have been unfolded into into deep learning models in \cite{lin2023target, liu2024bi, zhang2024joint, lin2024deep, nguyen2023joint}, achieving improved computational efficiency and performance. However, current unfolded algorithms are not applicable for addressing the fairness problem because of their problem-specific network structures. Furthermore, the current literature lacks unfolded learning models specifically designed for max-min fairness problems, which typically involve complex objective functions. 

In this paper, we fill this gap by jointly optimizing beamformers for both transmission and sensing, aiming to maximize the weighted sum of the minimum communications rate and sensing MI under per-antenna transmit power constraints. Given the non-smooth and non-convex nature of the problem, we first reformulate it into an equivalent but more tractable form and solve it using an alternating optimization (AO) procedure. Then, we propose a low-complexity yet efficient model-based ML method based on the AO framework. The proposed scheme retains the interpretability of conventional optimizers while achieving significantly better performance and shorter execution times. Furthermore, it scales effectively with the number of communication users, enabling seamless generalization to larger systems.

\section{System Model and Problem Formulation}\label{sec: system model}
{\bf JCAS System:}
We consider a downlink monostatic JCAS system, where a base station (BS) is equipped with $ \Nt $ transmit antennas and $ \Nr $ radar receive antennas. The same waveform is used for transmission and sensing \cite{liu2018toward,wang2024joint,fang2024beamforming}. The BS simultaneously transmits signals to $K$ single-antenna users for communications while probing $M$ targets. Letting $ \ssb=[s_1,\ldots, s_K]^\T \sim \mathcal{CN}(\mathbf 0,\Ib)$ be the vector of communications symbols, and  $ \Wb=[\wb_1,\ldots,\wb_K]\in\Cs^{\Nt\times K} $ be  the precoding matrix, the transmitted signal is expressed as
$ \xb= \Wb \ssb$. We assume an equal power allocation for all antennas, such that $\mathrm{diag}(\mathbf W\mathbf W^\H)\preceq  \frac{\Pt}{\Nt}\mathbf 1_{\Nt}$, where $\Pt$ is the  transmit power budget. 

The signal received by communications user $ k $ is given by
\begin{equation}\label{key}
\eqnsize
	y_k=\mathbf h_k^\H\mathbf w_ks_k+\mathbf h^\H_k\sum\nolimits_{j\neq k}^K\mathbf w_js_j +n_k, 
\end{equation}
where $ \mathbf h_k\in\mathbb C^{\Nt} $ denotes the channel between the BS and user $ k $, and $ n_k\sim\mathcal{CN}(0,\sigma_{\mathrm{c} k}^2)  $ represents additive white Gaussian noise (AWGN) with variance $\sigma_{\mathrm{c} k}^2$. We adopt the Rician fading model for $ \mathbf h_k, \forall k$. The SINR of the intended symbol at user $ k $ can be expressed as
\begin{equation}\label{key}
\eqnsize
	\gamma_{\mathrm{c} k}=\frac{|\mathbf h_k^\H\mathbf w_k|^2}{\sum_{j\neq k}^K|\mathbf h^\H_k\mathbf w_j|^2+\sigma_{\mathrm{c} k}^2}.
\end{equation} 

After transmission, the BS radar antennas receive echos reflected by the $M$ sensed targets as well as $C$ clutter scatterers around them. The received echo signal at the BS is written as
\begin{equation}\label{key}
\eqnsize
	\mathbf y_{\mathrm{s}}=\sum\nolimits_{m=1}^{M}\mathbf G_m\mathbf x+\sum\nolimits_{j=M+1}^{M+C}\mathbf G_j\mathbf x+\mathbf n_{\mathrm{s}},
\end{equation} 
where $ \mathbf G_i=\zeta_{\mathrm{s} i}\alpha_{\mathrm{s} i}\mathbf a_{\mathrm{r}}(\varphi_i)\mathbf a_{\mathrm{t}}^\H(\varphi_i), i=\{1,\ldots,M+C\} $, and $ \mathbf n_{\mathrm{s}}\sim \mathcal{CN}(0,\sigma_{\mathrm{s}}^2\mathbf I)  $ is AWGN. Here, we consider line-of-sight (LoS) channels between the BS and the target/clutters, which are assumed to be known for beamforming design \cite{liu2021cramer, song2023intelligent, nguyen2023joint}. In practice, LoS channel estimation can be performed efficiently using algorithms such as MUSIC. Parameters $ \zeta_{\mathrm{s} i}$, $\alpha_{\mathrm{s} i}$, and $\varphi_i $ represent the path loss, complex gain, and angle associated with target/clutter $i$, while $ \mathbf a_{\mathrm{r}}(\cdot)$ and $\mathbf a_{\mathrm{t}}(\cdot) $ are the normalized receive and transmit array steering vectors. The BS employs a receive combining matrix $ \mathbf F=[\mathbf f_1,\ldots,\mathbf f_M]\in\mathbb C^{\Nr\times M} $, such that the received signal associated with target $ m $ is given by
\begin{equation}\label{key}
\eqnsize
	 r_m=\mathbf f_m^\H\mathbf y_{\mathrm{s}}= \mathbf f_m^\H\mathbf G_m\mathbf x+\mathbf f_m^\H\sum\nolimits_{j\neq m}^{M+C}\mathbf G_j\mathbf x+\mathbf f_m^\H\mathbf n_{\mathrm{s}}.
\end{equation}
Thus, the SCNR can be expressed as
\begin{equation}\label{eq:SCNR}
\eqnsize
	\gamma_{\mathrm{s} m}=\frac{\|\mathbf f_m^\H\mathbf G_m\mathbf W\|^2 }{\sum_{j\neq m}^{M+C}\| \mathbf f_m^\H\mathbf G_j\mathbf W\|^2+\Nr\sigma_{\mathrm{s}}^2\|\mathbf f_m\|^2 }.
\end{equation}

\smallskip
{\bf Problem Formulation:}
We aim to jointly optimize the transmit precoding matrix $ \mathbf W $ and receive combining matrix $ \mathbf F $ to $1)$ ensure {\em fairness} among communications users and 
sensed targets; and $2)$ achieve a balanced tradeoff between communications and sensing performance. To achieve these goals, we employ the utility function $h(\Wb,\Fb)= \min_{k\in\mathcal K}\{ \log(1+\gamma_{\mathrm{c} k})\}+\delta \min_{m\in\mathcal M}\{ \log(1+\gamma_{\mathrm{s} m} )\}$, where $\log(1+\gamma_{\mathrm{c} k}) $ and  $\log(1+\gamma_{\mathrm{s} m} )$ represent the communications rate and sensing MI \cite{peng2024mutual,tang2018spectrally}, respectively, and $\delta\geq 0$ balances their performance tradeoff. The joint design problem is then formulated as 
\begin{equation}\label{P2}
\eqnsize
 \max_{ \mathbf W\in\mathcal S,\mathbf F}\,\,h(\Wb,\Fb),
\end{equation}
where $  \mathcal S = \{\mathbf W: \mathrm{diag}(\mathbf W\mathbf W^\H)\preceq  \frac{\Pt}{\Nt}\mathbf 1_{\Nt}\} $. The max-min objective guarantees fairness while the communications and sensing performance are balanced by adjusting $\delta$. However, problem \eqref{P2} is challenging, as the 
utility function includes the non-smooth point-wise minima in $h(\Wb,\Fb)$, non-convex fractional SINRs and SCNRs, and strongly coupled variables.
%

\section{Model-Based ML Optimizer}\label{sec: algorithm}
We herein propose a learning-aided optimizer for fairness JCAS beamforming. Specifically, we first formulate a surrogate objective to cope with the complex objective in \eqref{P2}. We identify an AO method suitable for  the resulting optimization. 
Then, we leverage model-based ML, and particularly deep unfolding~\cite{shlezinger2022model}, to enhance performance by converting the optimizer into a  ML model.

\smallskip
{\bf Surrogate Objective:}
Introducing the variables $ \mathbf z_{\mathrm{c}}=[z_{\mathrm{c}1},\ldots,z_{\mathrm{c} K}]^\T $ and $ \mathbf z_{\mathrm{s}}=[z_{\mathrm{s}1},\ldots,z_{\mathrm{s} M}]^\T $, we rewrite \eqref{P2} as

\begin{equation}\label{P2t1}
  \eqnsize
		\max_{ \mathbf W\in\mathcal S,\mathbf F}\,\, \min_{\substack{\mathbf z_c\in \mathcal Z_c,\\ \mathbf z_{\mathrm{s}}\in \mathcal Z_{\mathrm{s}}} }  \bigg\{ \sum\nolimits_{k=1}^{K} z_{\mathrm{c} k}r_{\mathrm{c} k}+\delta \sum\nolimits_{m=1}^{M} z_{\mathrm{s} m}r_{\mathrm{s} m}\bigg\},
\end{equation}
where $ \mathcal Z_c=\{\mathbf z_c:\mathbf z_c\succeq\mathbf 0, \mathbf 1_K^\T\mathbf z_c=1 \} $ and $ \mathcal Z_{\mathrm{s}}=\{\mathbf z_{\mathrm{s}}:\mathbf z_{\mathrm{s}}\succeq\mathbf 0, \mathbf 1_M^\T\mathbf z_{\mathrm{s}}=1 \} $ are compact convex simplices, and $r_{\mathrm{c} k}=\log(1+\gamma_{\mathrm{c} k}), r_{\mathrm{s} m}=\log(1+\gamma_{\mathrm{s} m} )$. As the optimal solution is located at the vertex of the simplices, solving~\eqref{P2t1} directly can lead to oscillation near the optimal point. To address this, we approximate $\{ \zb_c, \zb_{\mathrm{s}} \}$ by 
\begin{equation}\label{eq:Update z}
\eqnsize
     z_{\mathrm{c} k}\approx \frac{\exp(-\mu_c r_{\mathrm{c} k} )}{\sum_{j=1}^{K}\exp(-\mu r_{\mathrm{c}j}) }, \; 
 		z_{\mathrm{s} m}\approx \frac{\exp(-\mu_s r_{\mathrm{s} m} )}{\sum_{j=1}^{M}\exp(-\mu r_{\mathrm{s}j}) }, \forall m,k,
\end{equation}
where $ \mu_s $ and $\mu_c$ are continuous-valued parameters. These approximations become tight as $\mu_s, \mu_c,\rightarrow \infty$ \cite{xu2001smoothing}.
To deal with the non-convex objective, we introduce auxiliary variables $ \bm\xi_{\mathrm{c}}=[\xi_{\mathrm{c}1},\ldots,\xi_{\mathrm{c} K}]^\T\in\mathbb{R}^{K}$, $\bm\xi_{\mathrm{s}}=[\xi_{\mathrm{s}1},\ldots,\xi_{\mathrm{s} M}]^\T\in\mathbb{R}^{M} $, $\bm\theta_c=[\theta_{\mathrm{c}1},\ldots,\theta_{\mathrm{c} K} ]^\T\in\mathbb{C}^{K}$, and $\bm\Theta_{\mathrm{s}}=[\bm\theta_{\mathrm{s}1}^\H,\ldots,\bm\theta_{\mathrm{s} M}^\H ]^\H\in\mathbb{R}^{M\times K}$, and employ the quadratic transformation \cite[Theorem~1]{shen2018fractional} to obtain an equivalent but more tractable problem

\begin{equation}\label{pb:P2t2}
\eqnsize
		\max_{ \substack{\mathbf W\in\mathcal S,\\ \mathbf F,\bm\xi,\bm\Theta}}\,\, \min_{\substack{\mathbf z_c\in\mathcal Z_c,\\ \mathbf z_{\mathrm{s}} \in\mathcal Z_{\mathrm{s}}}} \bigg\{\sum\nolimits_{k=1}^{K} z_{\mathrm{c} k}O_{\mathrm{c} k}+\delta \sum\nolimits_{m=1}^{M} z_{\mathrm{s} m} O_{\mathrm{s} m} \bigg\},
\end{equation}
where $ \bm\xi=\{ \bm\xi_c,\bm\xi_{\mathrm{s}}\}, \bm\Theta=\{\bm\theta_c,\bm\Theta_{\mathrm{s}}\}$, 
\begin{align}
    O_{\mathrm{c} k} &= \log(1+\xi_{\mathrm{c} k}) + 2\sqrt{1+\xi_{\mathrm{c} k} }\Re\{\mathbf h_k^\H\mathbf w_k \theta_{\mathrm{c} k}^*\} -\xi_{\mathrm{c} k} \nonumber \\[-3pt]
    &\quad -|\theta_{\mathrm{c} k}|^2\left(\sum\nolimits_{j=1}^{K}|\hb_k^\H\mathbf w_j|^2+\sigma_{\mathrm{c} k}^2  \right), \label{eq_Ock}\\[-3pt]
    O_{\mathrm{s} m} &= \log(1+\xi_{\mathrm{s} m})+2\sqrt{1+\xi_{\mathrm{s} m}}\Re\{\mathbf f_m^\H\mathbf G_m\mathbf W\bm\theta_{\mathrm{s} m}^\H \} -\xi_{\mathrm{s} m} \nonumber\\[-3pt]
    &\hspace{-0.25cm} -\|\bm\theta_{\mathrm{s} m}\|^2\left(\sum\nolimits_{j=1}^{M+C}\| \mathbf f_m^\H\mathbf G_j\mathbf W\|^2+\Nr\sigma_{\mathrm{s}}^2\|\mathbf f_m\|^2  \right).
\end{align}

\vspace{-3mm}
\smallskip
{\bf AO Optimizer:}
 We can solve the problem \eqref{pb:P2t2} via AO. Specifically, the auxiliary variables $\{{\bm\xi},{\bm\Theta} \}$ are related to the design variables $\mathbf{F}$ and $\Wb$ via 
\begin{subequations}\label{Update auxi}
\eqnsize
	\begin{align}
		&\xi_{\mathrm{c} k}=\gamma_{\mathrm{c} k},\ \xi_{\mathrm{s} m}=\gamma_{\mathrm{s} m},\  \theta_{\mathrm{c} k}=\frac{\sqrt{1+\xi_{\mathrm{c} k}}\mathbf h_k^\H\mathbf w_k}{\sum_{j=1}^{K}|\mathbf h_k^\H\mathbf w_j|^2+\sigma_{\mathrm{c} k}^2 },\label{eq:xi and theta}\\
		&\bm\theta_{\mathrm{s} m}=\frac{\sqrt{1+\xi_{\mathrm{s} m} }\mathbf f_m^\H\mathbf G_m\mathbf W}{\sum_{j=1}^{M+C}\| \mathbf f_m^\H\mathbf G_j\mathbf W\|^2+\Nr\sigma_{\mathrm{s}}^2\|\mathbf f_m\|^2 }.\label{eq:vector theta}
	\end{align}
\end{subequations}
Hence, given $\{{\bm\xi},{\bm\Theta} \}$ and $\Wb$, $\Fb$ is updated by 
\begin{equation}\label{eq:f}
    \hspace{-2mm}\mathbf f_m\!=\! \frac{\sqrt{1+\xi_{\mathrm{s} m}}}{\|\bm\theta_{\mathrm{s} m}\|^2}\!\left(\sum\nolimits_{j=1}^{M+C}\!\!\mathbf G_j\mathbf W\mathbf W^\H\mathbf G_j^\H +\Nr\sigma_{\mathrm{s}}^2\mathbf I \right)^{-1}\!\!\!\!\!\!\!\! \mathbf G_m\mathbf W\bm\theta_{\mathrm{s} m}^\H.
\end{equation}
The proposed AO method thus operates in an iterative fashion. In each iteration, AO first uses $\mathbf{F}$ and $\Wb$ from the previous iteration to set the auxiliary variables \eqref{Update auxi}, which in turn are used to update $\Fb$ via \eqref{eq:f}. Then, $ \mathbf{W}$ is updated using projected gradient descent (PGD). Specifically, we define $\mathbf H=[\mathbf h_1,\ldots,\mathbf h_K ]$, $\mathbf 	X=\delta\sum\nolimits_{m=1}^Mz_{\mathrm{s} m}\sqrt{1+\xi_{\mathrm{s} m}}\bm \theta_{\mathrm{s} m}^\H\mathbf f_m^\H\mathbf G_m$, $\mathbf Y =\delta\sum\nolimits_{j=1}^{M+C} \mathbf G_j^\H \bigg(\sum\nolimits_{m=1}^Mz_{\mathrm{s} m}\|\bm\theta_{\mathrm{s} m}\|^2  \mathbf f_m\mathbf f_m^\H \bigg)\mathbf G_j$, and 
\begin{align}
    \mathbf\Sigma_1&=\mathrm{diag}\{z_{\mathrm{c}1}\sqrt{1+\xi_{\mathrm{c}1}}\theta_{\mathrm{c}1},\ldots,z_{\mathrm{c} k}\sqrt{1+\xi_{\mathrm{c} k}}\theta_{\mathrm{c} k} \}, \label{eq_sigma} \\
    \mathbf \Sigma_2 &=\mathrm{diag}\{z_{\mathrm{c}1}|\theta_{\mathrm{c}1}|^2,\ldots,z_{\mathrm{c} k}|\theta_{\mathrm{c} k}|^2 \}.  \label{eq_sigma2} 
\end{align}
With the other variables fixed, the subproblem with respect to $ \mathbf W $ is formulated as $\min_{ \mathbf W\in\mathcal S} g(\Wb)$,
where $g(\Wb) = \mathrm{tr}(\mathbf W\mathbf W^\H(\mathbf Y+\mathbf H\bm\Sigma_2\mathbf H^\H) )-2\Re\{\mathrm{tr}(\mathbf W (\mathbf X+\bm\Sigma_1^\H\mathbf H^\H) ) \}$ is convex with respect to $\Wb$. Hence, PGD can solve this problem. The gradient of $g(\Wb)$ is given by
\begin{equation}
\eqnsize
    \nabla_{\Wb}g = 2(\Yb+ \Hb\Sigmab_2\Hb^\H)\Wb -2(\Hb\Sigmab_1+\Xb^\H).
\end{equation}
Let $ \bm\Pi_{\mathcal S}(\cdot) $ denote the projection onto the set $ \mathcal S $: $\bm\Pi_{\mathcal S}(\Wb)= \sqrt{\frac{P_{\mathrm{t}}}{\Nt} }\diag(\Wb\Wb^\H)^{-\frac{1}{2}}\Wb$
where $ \diag(\cdot) $ is a diagonal matrix whose diagonal entries are taken from the argument. The PGD update is thus  
\begin{equation}\label{eq:PGD update rules}
\eqnsize
    \Wb\leftarrow \bm\Pi_{\mathcal S}\left(\tilde{\Wb}-\beta  \frac{ \nabla_{\Wb}g }{\| \nabla_{\Wb} g\|_F}\bigg|_{\Wb=\tilde{\Wb}} \right),
\end{equation}
where $\beta$ is the step size and $\tilde{\Wb}$ is the previous iterate of $\Wb$.

\smallskip
{\bf Unfolded Optimizer:} In the AO procedure, updating $\Wb$ requires a proper step size to ensure fast convergence. However, selecting an appropriate step size using techniques such as the backtracking search algorithm \cite{ma2024hybrid} introduces additional complexity. Furthermore, the parameters $\{ \mu_s, \mu_c\}$ also influence performance and typically require manual adjustment based on empirical experience. We overcome these difficulties by converting the AO into a discriminative ML model~\cite{shlezinger2022discriminative}, which can be viewed as a deep neural network (DNN). The DNN is trained to learn iteration-specific step sizes and $\{ \mu_s, \mu_c\}$ in an unsupervised manner, based on objective function $h(\Wb,\Fb)$.

\begin{figure}[t]
\small
    \centering	
     \includegraphics[width=0.4\textwidth]{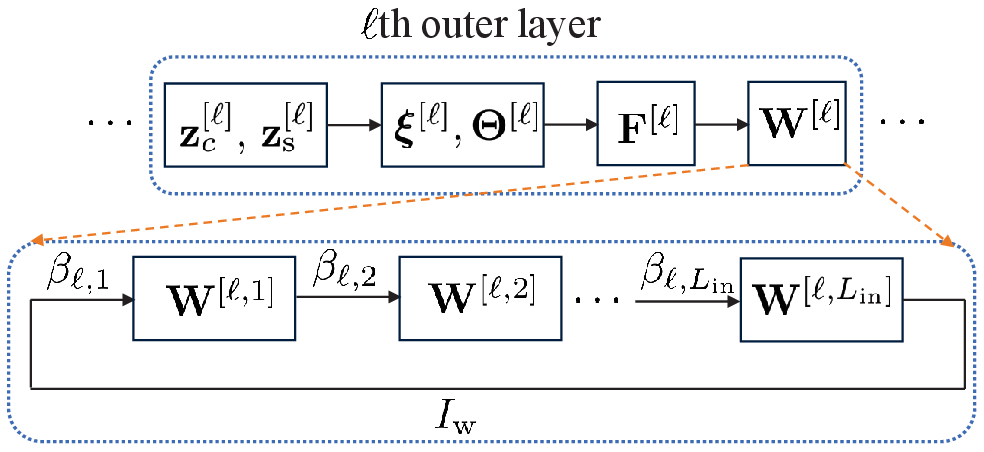}
     \vspace{-0.1cm}
        \caption{Proposed model-based DNN structure.}
        \label{fig:unfolding framework}
        \vspace{-2mm}
\end{figure}

The proposed DNN consists of $L_{\rm out}$ outer layers, and the structure of the $\ell$th outer layer is illustrated in Fig.~\ref{fig:unfolding framework}. In each outer layer, $\{\mathbf z_c,\mathbf z_{\mathrm{s}},\bm\xi,\bm\Theta,\Fb \}$ and $\Wb$ are updated successively, where the former ones have closed-form expressions given by \eqref{eq:Update z}, \eqref{Update auxi}, and \eqref{eq:f}. To efficiently obtain $\Wb$, we unfold the PGD algorithm into $L_{\rm in}$ inner layers. Specifically, with the trainable step size $\{\beta_{\ell,i}\}_{\ell=1,i=1}^{L_{\rm out},L_{\rm in}}$, $\Wb^{[\ell, i]}$ is updated by
\begin{equation}\label{eq:update W learning}
\eqnsize
  \Wb^{[\ell, i]} \! \leftarrow \! \bm\Pi_{\mathcal S}\!\left(\Wb^{[\ell, i-1]} \! -\!\beta_{\ell,i}  \frac{ \nabla_{\Wb}g }{\| \nabla_{\Wb} g\|_F}\bigg|_{\Wb=\Wb^{[\ell, i-1]} } \right).
\end{equation}
In each iteration,  $\Fb$ is obtained directly with closed-form solution  \eqref{eq:f}. In contrast, $\Wb$ is updated over unfolded DNN layers, and its update speed is restricted by the power constraint. Through simulations, we found that the update speed of $\Wb$ is much slower than that of $\Fb$, causing slow convergence of the overall algorithm. To overcome this, we propose updating $\Wb$ over $I_{\rm w}$ iterations of the $L_{\rm in}$ inner layers. Our simulations show that adjusting $I_{\rm w}$ can significantly accelerate the convergence, as will be demonstrated in Section~\ref{sec: numerical results}. 

The proposed AO-unfolded ML algorithm for solving~\eqref{P2} is outlined in Algorithm~\ref{alg:unfolding}. The input to the algorithm, denoted by $\bm d$, is the system configuration ${\bm d} = \{\Hb,\Gb,\Pt,\sigma_{\rm ck},\sigma_{\rm s} \}$, where $\Gb$ represents  $\{\Gb_m\}_{m=1}^{M+C}$. In step 1, we first set $\{\mu_s,\mu_c\}$ with predefined values and initialize $\Wb$ and $\Fb$ for each channel realization. Steps 2--13 represent the update of $\{\mathbf z_c,\mathbf z_{\mathrm{s}},\bm\xi,\bm\Theta, \Fb, \Wb\}$ with $L_{\rm out}$ outer layers, while in steps 5--12, the update of $\Wb$ is performed with $L_{\rm in}$ inner layers and $I_{\rm w}$ loops. We obtain $\{\Fb, \Wb\}$ from the output of the $L_{\rm out}$th outer layer, as in step 14.

The complexity of Algorithm~\ref{alg:unfolding} is dominated by the matrix multiplications required to update $\Fb$, $\Yb$, and $\Wb$. Computing $\Fb$ and $\Yb$ results in a complexity of $\Ocl(L_{\rm out}(M+C)(\Nt^2\Nr+\Nt\Nr^2))$. The complexity of updating $\Wb$ is $\Ocl(2L_{\rm out}L_{\rm in}I_{\rm w}K\Nt^2)$. Therefore, the overall complexity is approximately $L_{\rm out}\Ocl\big((M+C)(\Nt^2\Nr+\Nt\Nr^2)+2L_{\rm in}I_{\rm w}K\Nt^2\big)$.

\smallskip
{\bf Training:}
The trainable parameters of Algorithm~\ref{alg:unfolding} are thus $
{\bm \phi} =\{\mu_s,\mu_c,\{\beta_{\ell,i}\}_{\ell=1,i=1}^{L_{\rm out},L_{\rm in}}\}$. These are learned from a data set $\mathcal{D}$  comprised of $B$ potential JCAS settings, i.e., $\mathcal{D} = \{{\bm d}_b\}_{b=1}^B$. While the AO optimizer is derived from the proposed surrogate objective, the parameters of the unfolded AO are trained to maximize the original objective in \eqref{P2}. 

Specifically, letting $\Wb^{[\ell]}({\bm d};{\bm \phi}), \Fb^{[\ell]}({\bm d};{\bm \phi})$ denote the output of the $\ell$th iteration of Algorithm~\ref{alg:unfolding} with input ${\bm d}$ and parameters ${\bm \phi}$, the training loss used to guide the tuning of ${\bm \phi}$ from $\mathcal{D}$ is
\begin{equation}
\eqnsize
   \hspace{-2mm} \Lcl_{\mathcal{D}}( {\bm \phi})\! =\!-\frac{1}{B}\sum\limits_{b=1}^{B} \sum\limits_{\ell=1}^{L_{\rm out}} \lambda_l h(\Wb^{[\ell]}({\bm d}_b;{\bm \phi}), \Fb^{[\ell]}({\bm d}_b;{\bm \phi})),
\end{equation}
where $\lambda_l=\frac{1}{\ell}$ is the weight associated with the $\ell$th outer layer. The decreasing weight sequence prioritizes the initial layers for learning, allowing the solution to rapidly approach the optimal point in the beginning, and then gradually refined by the latter layers. In doing so, the overall convergence speed is slightly improved based on our simulations. 

\begin{algorithm}[t]
\small
\caption{Unfolded AO Optimizer}\label{alg:unfolding}
\LinesNumbered 
 \KwIn{${\bm d} = \{\Hb,\Gb,\Pt,\sigma_{\rm ck},\sigma_{\rm s}\}$}
\KwOut{$\Fb,\Wb$}
Initialization: Set $ \{\mu_s,\mu_c\}$ and generate $\{\Fb^{[0]},\Wb^{[0]}\}$.\\

\For{$\ell=1,\ldots,L_{\rm out}$}
{
    Update $ \mathbf z_c^{[\ell]}$, $ \mathbf z_{\mathrm{s}}^{[\ell]} $ based on \eqref{eq:Update z} with $\{\mu_s,\mu_c\}$.
  
  Update $ \bm\xi^{[\ell]},\bm\Theta^{[\ell]}$, and $\mathbf F^{[\ell]}$ based on \eqref{Update auxi} and \eqref{eq:f}.

    $\Wb^{[\ell, 0]}\leftarrow \Wb^{[\ell-1]}$.
    
    \For{$j=1,\ldots,I_{\rm w}$}{  
        
        \For{$i=1,\ldots,L_{\rm in}$}{
        Update $\Wb^{[\ell, i]}$ by \eqref{eq:update W learning}.
        }
        $\Wb^{[\ell, 0]}\leftarrow \Wb^{[\ell, L_{\rm in}]}$
    }
    $\Wb^{[\ell]} \leftarrow \Wb^{[\ell, L_{\rm in}]}$.
    
}

$\Fb=\Fb^{[L_{\rm out}]}, \Wb=\Wb^{[L_{\rm out}]}$.

\end{algorithm}

\section{Numerical Results}\label{sec: numerical results}
 In this section, we evaluate the performance of the proposed model-based ML algorithm\footnote{The  code source is available online at \url{https://github.com/WillysMa/JCAS_BF_Design_MaxMin}.}. Throughout the simulations, we set $\Nt\!=\!\Nr\!=\!16$, $M\!=\!C\!=\!2$, and $K\!=\!4$ for training and testing unless otherwise stated. We set Rician factor of $3$ dB to generate $\mathbf{h}_k, \forall k$. The path loss is modeled as $\zeta_{\mathrm{x}} = \zeta_0 d_{\mathrm{x}}^{\epsilon_{\mathrm{x}}}$, where $\zeta_0 = -30$ dB is the reference path loss at $1$ m, and $(\cdot)_\mathrm{x}$ with $\mathrm{x} \in \{\mathrm{c}, \mathrm{s}\}$ represent the parameters for communications and sensing channels, respectively. Accordingly, we assume the path loss exponents $\epsilon_\mathrm{c} = 3$, $\epsilon_\mathrm{c} = 2$ and distances from the BS to the $k$th user and $m$th target as $d_{\mathrm{c}, k}=100+20\eta_{\mathrm{c}, k}$ and $d_{\mathrm{s}, m}=10+2\eta_{\mathrm{s}, m}$, where $\eta_{ck},\eta_{sm}\sim\mathcal{N}(0,1)$. The transmit power is set to achieve a signal-to-noise ratio (SNR) of $ \frac{P_{\rm t}}{\sigma_{\rm s}}=20$~dB, with $\sigma_{\rm s}=\sigma_{\rm ck}=-80$~dBm.
 
 The sizes of the training and testing sets are $500$ and $100$, respectively, using the general Rician fading model for communications channels. We train over over $30$ epochs with batch size of $32$. For initialization, we set $\beta_{\ell,i}=0.01,\; \forall \ell,i$, and initialize $\Wb$ based on the maximum ratio transmission method, i.e.,$\Wb^{[0]}=\bm\Pi_{\mathcal S}(\Hb)$. Furthermore, we set $\mu_s= \mu_c=10$ based on empirical simulations, which ensures the convergence of both the proposed method and the compared scheme in \cite{fang2024beamforming}. We set $\Fb^{[0]}$ by maximizing the SCNR in \eqref{eq:SCNR} for the $M$ targets, which is a generalized eigenvalue problem whose solution is established in the literature \cite{ghojogh2019eigenvalue}. For all results reported below, we set $L_{\rm out}=150, L_{\rm in}=3, I_{\rm w}=2$ for training. All results are obtained by averaging over $100$ channel realizations.
\begin{figure}[t]
\small
\vspace{-0.25cm}
    \centering
    \hspace{-5mm}
    \subfigure[$\delta=1$.]
    {\label{fig:convergence1}\includegraphics[width=0.24\textwidth]{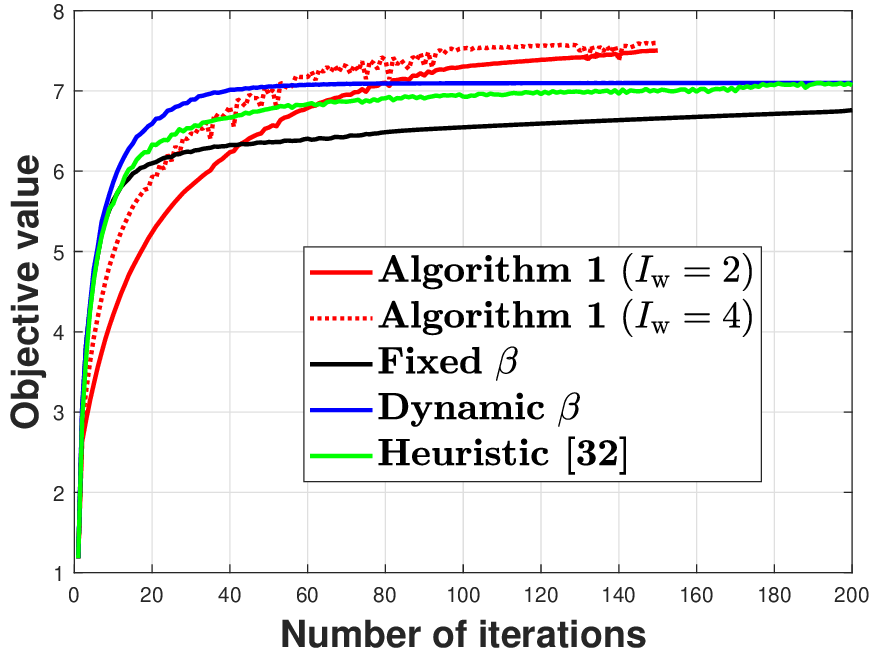}}
     \hspace{-3mm} 
    \subfigure[$\delta=10$.]
    {\label{fig:convergence10} \includegraphics[width=0.24\textwidth]{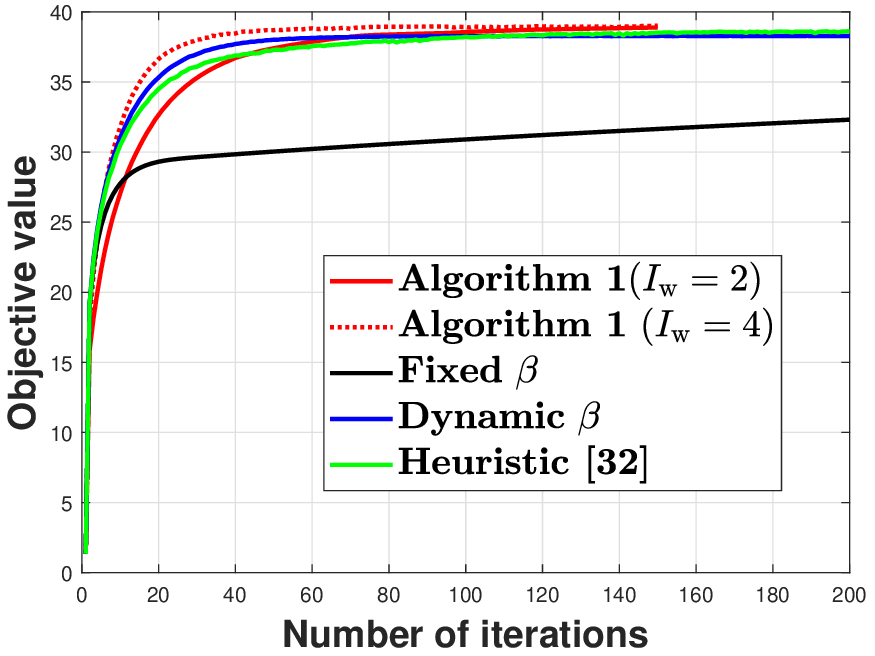}}
    \vspace{-3mm}
    \caption{Convergence of Algorithm \ref{alg:unfolding} with $\delta = \{1, 10\}$.}
    \label{fig:convergence} 
        \vspace{-2mm}
\end{figure}

  \begin{figure}[t]
\small
    \centering
    \hspace{-5mm} 
        \subfigure[Min. SINR and SCNR for $\delta=1$.]
    {\label{fig:SINR} \includegraphics[width=0.24\textwidth]{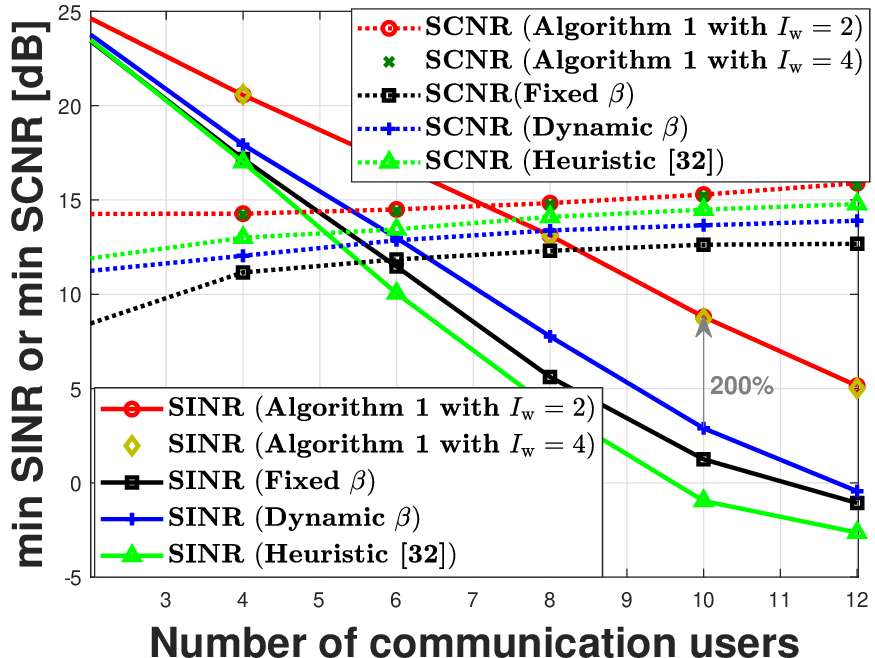}}
        \hspace{-2mm}
    \subfigure[Comm--sensing tradeoff region.]
    {\label{fig:tradeoff} \includegraphics[width=0.24\textwidth]{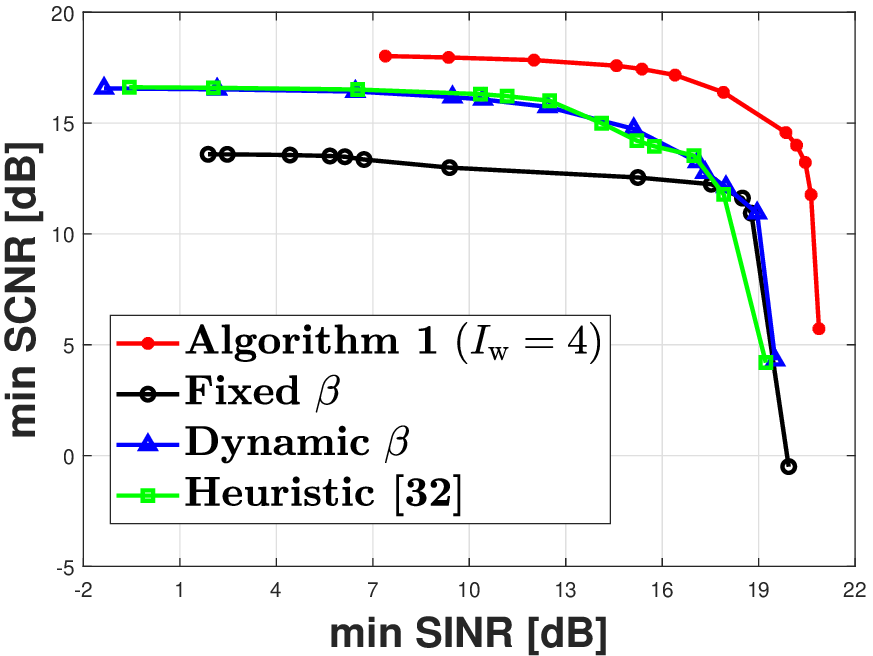}}
    \vspace{-2mm}
    \caption{Minimum SINR and minimum SCNR and their tradeoff.}
    \label{fig:performance} 
        \vspace{-2mm}
\end{figure}

\begin{table}[t]
\small
\renewcommand\arraystretch{1.1}
\centering
\caption{Average run time ([s]) of the considered algorithms.} \label{tb:run time}
    \begin{tabular}{c|c c c c} 
    \hline 
      \textbf{Algorithm}  &$K=6$ & $K=8$ & $K=10$ & $K=12$\\
          \hline    
          \hline
      Fixed $\beta$   &$0.4482$ & $0.9897$&$1.420$ &$1.736$ \\
      \hline
           Heuristic~\cite{fang2024beamforming} &$0.3504$ & $0.4352$ &$0.4740$ &$0.4943$\\
       \hline 
    Dynamic $\beta$ &$0.1985$ & $0.2892$&$0.3540$ &$0.2775$ \\
    \hline 

    Algorithm~\ref{alg:unfolding} ($I_{\rm w}=2$) &$0.2249$ & $0.2327$&$0.2308$ & $0.2313$  \\
        \hline 
    Algorithm~\ref{alg:unfolding} ($I_{\rm w}=4$)&$\bm{0.1730}$ & $\bm{0.1704}$&$\bm{0.1774}$ &$\bm{0.1719}$ \\
    \hline 
    \end{tabular}
\end{table}

 For comparison, we consider AO optimizers employing either fixed (``Fixed $\beta$'') or dynamic (``Dynamic $\beta$'') step sizes to update $\Wb$. For the former, the step size is set to $0.01$, which is the same as the initial value for $\beta_{\ell,i}\; \forall \ell,i$. For the latter, the step size for each $\Wb$ update is found by the backtracking line search method. The steps of the two algorithms are the same as in Algorithm~\ref{alg:unfolding} except we set $I_{\rm w}=1$. 
 We also compare with the heuristic method for updating $\Wb$ in~\cite{fang2024beamforming}.
 
In Figs.~\ref{fig:convergence1} and \ref{fig:convergence10}, we show the convergence of the considered schemes with $\delta \in \{1,10\}$ and $I_{\rm w} = \{2,4\}$. It is observed that the proposed algorithm converges to the largest objective values with fewer iterations compared to the benchmark schemes. The gain is more significant for smaller $\delta$. Furthermore, increasing $I_{\rm w}$ accelerates the convergence and leads to a higher objective value of the proposed method. However, a large $I_{\rm w}$ may cause oscillation near the local optimal point, as can be observed in Fig.~\ref{fig:convergence1} for $I_{\rm w} = 4$.  

Fig.~\ref{fig:SINR} shows the minimum SINRs and SCNRs for $K = \{2,4,\ldots,12\}$ and $\delta=1$. As $K$ increases, inter-user interference becomes more significant. Thus, the minimum SINR of all the schemes considered decreases, contributing less to the objective function $h(\Wb,\Fb)$ in \eqref{P2}. As a result, the minimum SCNR increases slightly with $K$. We note here that the proposed unfolding model is trained with the dataset obtained for $K=4$. However, it generalizes well for various values of $K$ during online inference because its model structure is independent of $K$. Among the compared schemes, the proposed method achieves the largest minimum SINR/SCNR, with its gain being more significant for the communications functions. With $K=10$, the SINR improves by up to $200\%$ compared to the scheme using backtracking line search. We further show the tradeoff between the minimum SINR and SCNR of the considered schemes for $\delta\in [0, 100]$ in Fig.~\ref{fig:tradeoff}. It is seen that the proposed scheme achieves the best tradeoff between the minimum SINR and minimum SCNR.

 In Table~\ref{tb:run time}, we show the average run time of the considered algorithms for $\delta=1$. All the compared schemes are performed on the same platform. Based on the results in Fig.\ \ref{fig:convergence1}, we set $L_{\rm out}= \{150, 100\}$ for $I_{\rm w}=\{2,4\}$, respectively, for Algorithm~\ref{alg:unfolding}. Meanwhile, the benchmark schemes are set to run until convergence. It is observed from Table~\ref{tb:run time} that the proposed scheme with $I_{\rm w}=4$ performs the fastest, and its run time remains almost constant when $K$ increases. Furthermore, the reduction in run time of the proposed scheme is more significant with larger $K$. For example, when $K = 10$, Algorithm~\ref{alg:unfolding} with $I_{\rm w}=4$ achieves a run time reduction of $67\%$, $75\%$, and $92\%$ compared to the dynamic step size strategy, the heuristic method, and the fixed step size strategy, respectively. Note that this reduction is achieved while better performance is guaranteed, as seen from Fig.~\ref{fig:SINR}. The superior performance of Algorithm~1 over the conventional optimization-based approaches is due primarily to the fact that the step sizes are learned from the data, which enhances convergence. 
 

\vspace{-5pt}
\section{Conclusions}
This paper proposed an efficient beamforming design for multi-user multi-target JCAS systems, aiming to ensure fairness and balance between communications and sensing performances. We jointly optimized the transmit and receive beamformers to maximize the weighted sum of the minimum communications rate and sensing MI. To address this challenging problem, we proposed a model-based ML algorithm. Numerical results demonstrate that our algorithm scales well with the number of communications users while delivering better performance with shorter run time compared to conventional optimization-based methods.
	

	
\clearpage
    \vspace{-1cm}
\let\OLDthebibliography\thebibliography
\renewcommand\thebibliography[1]{
    \OLDthebibliography{#1}
    \setlength{\parskip}{0pt}
    \setlength{\itemsep}{0pt plus 0.0ex}
}
\bibliographystyle{IEEEbib} 
\bibliography{IEEEabrv,reference}
\end{document}